\documentclass[aps,prl,twocolumn,superscriptaddress,
               amsmath]{revtex4}

\usepackage{graphicx}% Include figure files
\usepackage{dcolumn}% Align table columns on decimal point
\usepackage{bm}% bold math

\begin{document}

\title{
Two-Dimensional Quantum XY Model with Ring Exchange and 
External Field
}

\author{R.~G.~Melko}
%\email{rgmelko@physics.ucsb.edu}
\affiliation{Department of Physics, University of California, Santa
Barbara,
California 93106}

\author{A.~W.~Sandvik}
%\email{asandvik@abo.fi}
\affiliation{Department of Physics, {\AA}bo Akademi University,
Porthansgatan 3, FIN-20500 Turku, Finland}

\author{D.~J.~Scalapino}
%\email{djs@vulcan2.physics.ucsb.edu}
\affiliation{Department of Physics, University of California, Santa
Barbara,
California 93106}

\date{\today}

\begin{abstract}
We present the zero-temperature phase diagram of a square lattice quantum 
spin 1/2 XY model with four-site ring exchange in a uniform external magnetic
field. Using quantum Monte Carlo techniques, we identify various quantum phase
transitions between the XY-order, striped or valence bond solid, staggered 
N\'eel antiferromagnet and fully polarized ground states of the model. We find
no evidence for a quantum spin liquid phase.

\end{abstract}

\maketitle

Studies of two-dimensional spin-1/2 quantum magnet and boson
models have provided insight into novel quantum phases and quantum
critical points \cite{Gen1}.  Recently, interest has focused on models
which have multi-site ring exchange \cite{Ring1,Arun,LMGH}.  
The ring exchange interaction, either alone or in competition with 
the usual spin or boson near-neighbor exchange, has been shown to 
promote a variety of exotic quantum ground states \cite{Arun}, including in 
some cases a spin-liquid state \cite{LMGH}.
Of particular importance is the class of two-dimensional model Hamiltonians 
that contain quantum spin-1/2 or boson operators interacting with ring
exchange that can be simulated using quantum 
Monte Carlo (QMC) techniques without a negative sign problem.
With modern algorithms, such models can be studied numerically on large 
lattices without approximation, providing a laboratory for surveying the 
critical behavior that separates various quantum phases.

One important model in this respect is the easy-plane $J$-$K$ model \cite{AD1}
that has quantum $S=1/2$ spins on a square lattice with a near-neighbor
exchange $J$ and a four-site ring exchange $K$.  
This Hamiltonian is partially motivated by the undoped cuprate materials
\cite{CupRing},
where ring-exchange processes are believed to contribute to experimental
signatures beyond those explained by the near-neighbor Heisenberg model.
The two-parameter $J$-$K$ model, despite its simplicity, 
displays a surprisingly rich and complex phase diagram \cite{AD1,AD2}, 
with three
distinct zero-temperature phases.  These are an XY-ordered or superfluid 
phase for
large $J$, a staggered N\'eel or boson charge density wave (CDW) phase for large
$K$, and a striped or valence bond solid (VBS) phase for intermediate
$K/J$.  
The zero-temperature phase transition between the VBS and N\'eel phases
is first order, however previous numerical results \cite{AD1,AD2} 
indicate the existence of
a continuous quantum critical point (QCP) at the zero-temperature 
superfluid-VBS boundary.  
%This QCP has very recently attracted much theoretical attention, as it
%is now believed to exhibit novel complex behavior of broad interest in the 
%field of quantum phase transitions \cite{LeonMatt}.

\begin{figure}[ht]  
\begin{center}
\includegraphics[height=6.8cm]{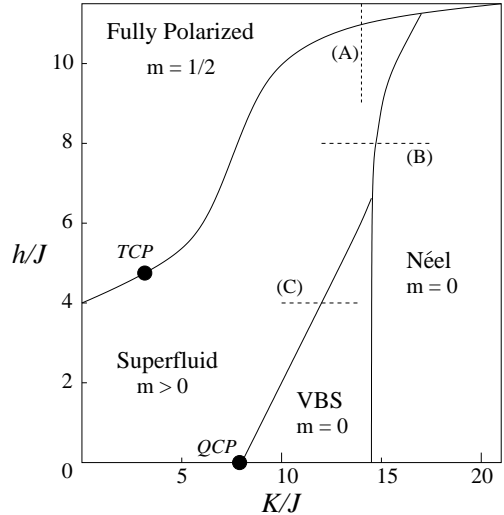}
\caption{
The schematic zero-temperature phase diagram of the easy-plane $J$-$K$-$h$ 
model.  Phase boundaries are drawn as solid lines.  Dashed lines indicate cuts
along which we have examined the transitions between the various phases,
as discussed in the text.
}
\label{PhaseD}
\end{center}
\end{figure}

The question naturally arises as to the behavior of the easy-plane
$J$-$K$ model under the influence of a magnetic field.
This is interesting both as a study of the 
evolution of the QCP, as well as the behavior of the 
ground state phases away from half-filling.
Using stochastic series expansion (SSE) QMC, we present here the basic
features of the zero-temperature phase diagram of the easy-plane model 
described by the Hamiltonian
\begin{equation}
H = -J \sum_{\left< ij \right>}B_{ij} - K \sum_{\left< ijkl
\right>}P_{ijkl} - h \sum_i S_i^z,
\label{KJHhamil}
\end{equation}
where $S_i^z$ is the $z$ component of a quantum spin-1/2, 
$B_{ij} = S_i^+ S_j^- + S_i^- S_j^+$ is a near-neighbor exchange, 
and $P_{ijkl} = S_i^+ S_j^- S_k^+
S_l^- + S_i^- S_j^+ S_k^- S_l^+$ generates a four-site ring exchange.
Here, $\left< ij \right>$ denotes a pair of nearest-neighbor sites and 
$\left< ijkl \right>$ are sites on the corners of a square plaquette
on the $L \times L$ lattice.
For $K=0$, this is the standard XY-model in a uniform magnetic field, or 
alternatively hard-core bosons with a chemical potential.  For $h=0$, this model 
is in a XY-ordered or superfluid phase for temperatures less than the 
Kosterlitz-Thouless 
transition temperature of $T_{KT}/J\approx0.68$ \cite{HaradaMelko}.
With the application of a uniform magnetic field the average magnetization
$m= \left<{S^z}\right>$ of the XY-superfluid increases from zero ($m=0$) until 
it saturates into a fully spin polarized ($m=1/2$) state at $h/J=4$ \cite{Bern1}.   
For $J=0$ and $h=0$, the ground state of the system is N\'eel antiferromagnetic
order \cite{Arun,AD1}.
For $h=0$, it was found \cite{AD1,AD2} that an intermediate VBS
phase exists for $7.9 \lesssim K/J \lesssim 14.5$, in which the expectation 
value $\left<{P_{ijkl}}\right>$ alternates in strength with a period of two
lattice spacings in one of the lattice directions, suggesting the term
``striped'' order.

To study the effect of the uniform magnetic field $h$ on
the ground state properties of the easy-plane $J$-$K$ model,
we use the SSE quantum Monte Carlo simulation method \cite{SSE1} that
was previously applied to the $h=0$, $J$-$K$ model \cite{AD1,AD2}.
In order to implement the SSE method, the operators in the
Hamiltonian (\ref{KJHhamil}) are represented as four-spin {\it
plaquette} operators.  Diagonal operators involving $h$ terms 
are added to or removed from the SSE basis state expansion using a
simple Metropolis probability algorithm.  Off-diagonal ($J$ or $K$ term)
operators are sampled using the {\it directed-loop} algorithm
\cite{SSE1,AD2}, which becomes increasingly important for simulation
efficiency with increasing magnetic field strength.  The directed-loop
equations \cite{SSE1} for the $J$-$K$-$h$ model are only slightly 
more complicated than for the pure $J$-$K$ model \cite{AD2}, 
and are presented elsewhere \cite{RAD1}.
The QMC algorithms were tested on $L=4$ lattice sizes against exact
diagonalization results and previous QMC simulations on the pure XY and
$J$-$K$ models.
In this paper, simulations were carried out on square lattices of linear 
dimension $L$ (number of spins $N=L^2$), at temperatures $T=1/\beta$ low
enough to ensure convergence into the ground state.

A variety of physical observables of direct relevance to the ground
states of the model are accessible through the SSE method. 
It is straightforward to calculate the internal energy \cite{SSE1}
since its statistical estimator is just the number $n$ of plaquette 
operators in the SSE basis-expansion operator sequence
multiplied by $T$: $E=-\left<n\right>/\beta$.
The spin stiffness (or superfluid density in the boson representation)
is defined in terms of the energy response to a twist $\phi$ in the
periodic boundary of the lattice by
\begin{equation}
\rho_s = \frac{\partial^2 E}{\partial \phi^2},
\end{equation}
and is directly estimated using the winding number fluctuations in the
SSE simulation \cite{CeperlyAWS}.  In addition we calculate
the plaquette structure factor
\begin{equation}
S_p(q_x,q_y) = \frac{1}{L^2}\sum\limits_{a,b}
{\rm e}^{i({\bf r}_a-{\bf r}_b)\cdot {\bf q}}
\langle P_{a_1a_2a_3a_4}P_{b_1b_2b_3b_4} \rangle .
\end{equation}
Here, $a_1,\ldots,a_4$ are the sites belonging to plaquette $a$,
located at ${\bf r}_a$.  In the VBS phase, the square of the 
magnitude of the order parameter per site is
$\left<{M_P}\right>^2=[S_p(\pi,0)+S_p(0,\pi)]/2L^2$.
Similarly, the square of the order parameter 
$\left<{M_S}\right>$ of the N\'eel ordered phase is 
obtained from the $S^z$ structure factor
\begin{equation}
S_s(q_x,q_y) = \frac{1}{L^2}\sum\limits_{j,k}
{\rm e}^{i({\bf r}_j-{\bf r}_k)\cdot {\bf q}}
\langle S^z_jS^z_k \rangle ,
\label{SSstrF}
\end{equation}
with $\left<{M_S}\right>^2=S_s(\pi,\pi)/L^2$.
Here, $j$ and $k$ are lattice sites located at lattice 
coordinate ${\bf r}_j$.  The quantities 
$\left<{M_P}\right>^2$ and $\left<{M_S}\right>^2$
are expected to decrease as
$1/L^2$ (signifying short-range correlations) in phases without the
respective order, but tend to a finite value for large $L$ in phases
where long-range order occurs.
\begin{figure}[ht]
\begin{center}
\includegraphics[height=6.8cm]{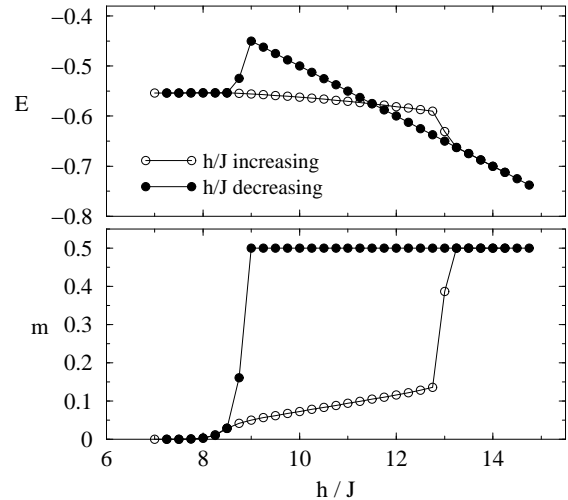}
\caption{
The ground-state energy ($E$) and magnetization ($m$) of an $L=8$ system 
along cut (A) in the phase diagram Fig.~\ref{PhaseD}.
This set of simulations was performed with parameters $K/J=14$ and
$\beta J=320$.  The hysteresis effects were obtained by systematically increasing
and then decreasing $K/J$ in steps, with system configurations stored
at the end of one $K/J$ step and used to begin the next.
}
\label{Em}
\end{center}
\end{figure}

By directly observing the behavior of the spin stiffness (superfluid 
density) and the VBS and N\'eel order parameters, we are able to map out the
phase boundaries of the $J$-$K$-$h$ model as illustrated in
Fig.~\ref{PhaseD}.
In general, we find no persistent regions of quantum disorder (i.e. a spin
liquid state) in the vicinity of the $h=0$ quantum critical point.
Rather, the QCP appears to evolve smoothly into a quantum phase
transition between the superfluid and VBS regions for 
$0 \le h \lesssim 6$.  The
$J$-$K$-$h$ model also exhibits a direct superfluid to N\'eel order
transition for $6 \lesssim h \lesssim 11$, a feature not contained in the 
$h=0$ phase
diagram.  Finally, for large $h$, the model finds a fully 
polarized spin state with $m=1/2$.  This latter phase
transition is strongly first order for $K/J \gtrsim 5$,
displaying pronounced metastability and hysteresis effects in the
simulation (see Fig.~\ref{Em}).
Renormalization Group treatments of two-dimensional bosons
\cite{FishSub}, as well as spin-wave corrected mean-field theory and
simulations of a hard-core boson Hamiltonian \cite{Bern1} indicate that
at $K=0$, the pure XY model exhibits a continuous transition to the
fully polarized state at $h=4J$.
This suggests that a tricritical point (TCP in Fig.~\ref{PhaseD}) exists on the 
phase boundary somewhere between $0<K/J \lesssim 5$, above which the 
transition to the fully polarized state becomes first order.

\begin{figure}[ht]
\begin{center}
\includegraphics[height=8cm]{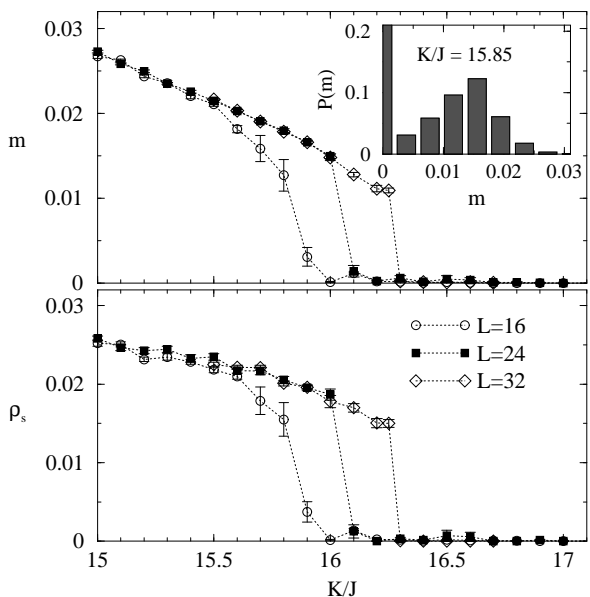}
\caption{
Magnetization ($m$) and spin stiffness ($\rho_s$) of the superfluid-N\'eel
transition, along cut (B) in the phase diagram Fig.~\ref{PhaseD}.
Model parameters are $h/J=8$, $\beta J=320$ for $L=16$ and 
$\beta J=400$ for the larger lattice sizes.  The inset shows a
double-peaked magnetization probability histogram $P(m)$ representing 
$5~\times~10^5$ Monte Carlo steps at a point on the $L=16$ data curve in the 
transition region.  
The lower peak is not Gaussian in shape, as the system is attracted to
zero magnetization (the half-filled) state.}
\label{SFCH}
\end{center}
\end{figure}

The energy crossover and density hysteresis of Fig.~\ref{Em} provide one 
indicator of a first-order transition.  Alternatively, one
may look for an abrupt discontinuity in the order parameter (for large
system sizes) or for double-peaked probability histograms for data 
in the transition region. To illustrate this
we turn now to a detailed set of simulation results for the
superfluid-N\'eel phase boundary along cut (B) in Fig.~\ref{PhaseD}.
As illustrated in Fig.~\ref{SFCH}, the boson and superfluid
densities develop significant discontinuities for larger systems 
as the phase boundary is traversed. This abrupt discontinuity does not appear
for $L<20$, illustrating that the transition is caused by an avoiding level
crossing and that large lattices sizes are necessary to quantify
the behavior of this model. The first-order nature is apparent in 
double-peaked magnetization histograms, that were observed for data in 
the ``discontinuity'' regions for $L=16$, indicating a phase 
coexistence. For $K/J \gtrsim 16.3$, the spin spin structure factor 
(Eq.~(\ref{SSstrF}))
develops Bragg peaks at $(\pi,\pi)$ (not illustrated), indicating N\'eel 
order.  
It is interesting to note that a similar phase transition
between a superfluid and a $(\pi,\pi)$ staggered solid is
found in hardcore boson Hubbard models with nearest and next-nearest
neighbor repulsion \cite{Schmid1,Hebert}.

\begin{figure}[ht]
\begin{center}
\includegraphics[height=9.3cm]{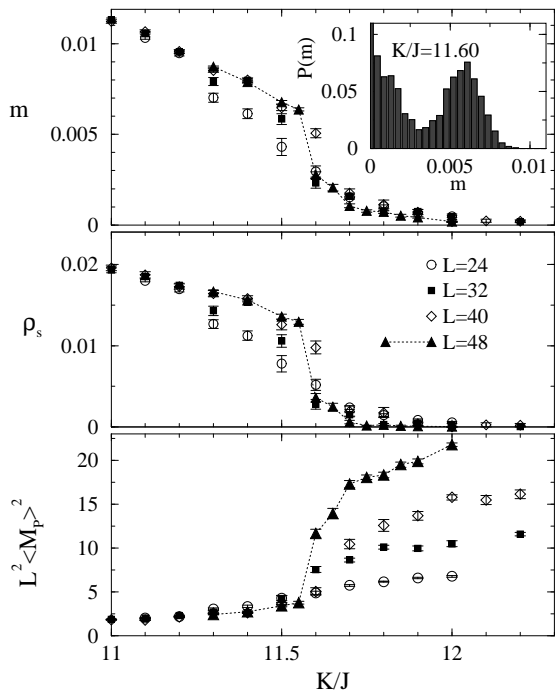}
\caption{
Details of the magnetization, spin stiffness and the VBS structure factor
of the superfluid-VBS transition, along cut (C) in the phase diagram
Fig.~\ref{PhaseD}.
Model parameters are $h/J=4$, $\beta J=320$ for $L=24$ and
$\beta J=400$ for the larger lattice sizes.
The inset shows a double-peaked magnetization probability histogram
$P(m)$ representing $3.5~\times~10^5$ Monte Carlo steps at a point on the
$L=48$ data curve in the transition region.
}
\label{SFstr}
\end{center}
\end{figure}

Finally we examine the XY superfluid-VBS transition along cut (C).
As illustrated in Fig.~\ref{SFstr}, simulation data
for system size $L=24$ does not display an obvious sharp discontinuity as in 
Fig.~\ref{SFCH}.  However, the presence of a small discontinuity in $m$ and 
$\rho_s$ for $L=32$ to 48 is suggested by the data.
The inset of Fig.~\ref{SFstr} displays a double-peaked magnetization
probability histogram in the transition region, which indicates the presence 
of a first-order phase coexistence.  
%It is interesting to note that the lower peak of 
%this histogam is much broader than the lower peak of the histogram in
%the inset of Fig.~\ref{SFCH}, even though in Fig.~\ref{SFstr} the
%smaller value for $h$ produces a much more subtle phase separation.
%The double-peaked density histogram of Fig.~\ref{SFstr} (inset)
This clearly
precludes the existence of a continuous quantum phase transition, at
least for the field value $h/J = 4$ that was studied in cut (C) (see 
Fig.~\ref{PhaseD}).  The most immediate
conclusion to draw is that the superfluid-VBS phase transition is
weakly first order, either along its entirety (excluding the $h=0$ QCP), 
or up to a tricritical point at a field $0 < h < 4$.
In this case, the difficulty in seeing a large discontinuity in the 
superfluid density or plaquette structure factor is due to the small $h$ value and
the closeness of the magnetization to zero.  The persistence
of a small region of superfluid density in apparent coexistence with
a finite VBS order parameter (for example, the two data points for
$L=48$, $K/J = 11.60$ and 11.65 in Fig.~\ref{SFstr}) 
is due to the first-order metastability 
between the superfluid phase and the VBS phase that is obscured by
statistical averaging.  
As a check, we observed the Monte Carlo time correlation
between $S_p(\pi,0)$, $S_p(0,\pi)$, and the superfluid density 
$\rho_s$ in the $x$ and $y$-directions at these points.
In fact, we find that $\rho_s$ show no
preference for the $x$ or $y$ directions when stripe 
order is present. Rather, both $\rho_s^y$ and $\rho_s^x$
show very strong anti-correlations whenever $\left<{M_P}\right>^2$
develops Bragg peaks.

In summary, using SSE QMC techniques, we have determined the ground
state phase diagram (Fig.~\ref{PhaseD}) of the easy-plane $J$-$K$-$h$
model.
In addition to the XY superfluid, VBS, and N\'eel ordered phases observed 
for $h=0$ \cite{AD1}, we observe a large region of 
fully polarized order, which dominates the phase diagram for large $h$.
The phase transition to the polarized state is continuous at 
small $K/J$ and strongly
first order for large $K/J$, suggesting the existence of a tricritical
point somewhere on the phase boundary for intermediate $K/J$.
Two other phase transitions were studied in detail, the superfluid-N\'eel
and superfluid-VBS transitions. Both were first order for the parameter
values investigated in detail here.

As indicated by our data, the $J$-$K$-$h$ model does not appear to
support a region of superfluid-VBS coexistence (i.e. a supersolid), 
which is observed near a similar transition between a superfluid and 
$(\pi,0)$ striped solid phase in a hardcore boson Hubbard model 
\cite{Hebert,HubBosRG}.
No additional ordered phases were observed in this model,
in particular, incommensurate VBS stripes (or striped order away from 
half filling) which would have been indicated by Bragg peaks in the 
{\bf q}-dependent structure factor $S_p(q_x,q_y)$ away from $(\pi,0)$.  
%This is significant, as it suggests that XY ring-exchange does not 
%promote a true charge stripe ordered phase (CSO), and consequently no
%regions of coexisting superfluid-CSO order like those investigated in 
%reference to the cuprates \cite{Kiv1}. 

In the context of the $h=0$ superfluid-VBS transition at $T=0$
\cite{AD1,AD2},
%the exact nature of the $h>0$ quantum phase transition is not essential;
%i.e. 
the existence of a continuous QCP does not require a
continuous phase transition to develop smoothly as $h$ is increased from
zero.  Conversely, the existence of a true continuous phase transition would
provide additional supporting evidence for the existence of the
$h=0$ QCP \cite{RogNote}, as well as a further region in which to explore the
nature of the critical behavior associated with the transition from 
XY superfluid to VBS order.
Ultimately, one would like to determine whether this QCP is an example 
of the ``deconfined'' quantum criticality recently discussed by
Senthil {\it et al.} \cite{LeonMatt}.

%and another model to calculate dynamical scaling 
%exponents \cite{AD2}.  
%Such studies will prove to be invaluable as references for theorists,
%who are currently examining the $h=0$ QCP as an example of a 
%special ``deconfined'' QCP with behavior that lies outside the usual
%Landau-Ginzburg-Wilson framework of phase transitions driven by fluctuating  
%order parameters \cite{LeonMatt}.
%However, differentiating numerically between a continuous or a weakly first-order
%superfluid-VBS phase transition for fields less than $h/J = 4$ would be 
%exceeding difficult using the techniques illustrated in this study, requiring 
%lattice sizes of $10^4$ sites or more.

Finally, the inability of any significant region of a spin-liquid phase
to develop in the vicinity of the QCP motivates further searches on related 
models.  Of particular interest is the square lattice $J$-$K$ ring model in
a {\em staggered} magnetic field, which could conceivable destabilize the 
superfluid or VBS order near the QCP and promote the development of an 
extended region of disorder.
Work on this model is in progress \cite{RAD1}.

The authors would like to thank L. Balents and M.~P.~A.~Fisher for insightful 
discussions and a critical reading of the manuscript. This work was supported
by the Department of Energy, Grant No.~DE-FG02-03ER46048 (DJS),
and by The Academy of Finland, project No.~26175 (AWS). Supercomputer 
time was provided by NCSA under grant number DMR020029N. AWS would like to
thank the Department of Physics at UCSB for hospitality and support during 
a visit.


\begin{thebibliography}{99}
\bibitem{RogNote} However, differentiating numerically between a 
continuous or a weakly first-order
superfluid-VBS phase transition for fields less than $h/J = 4$ would be
exceeding difficult using the techniques illustrated in this study, requiring
lattice sizes of $10^4$ sites or more.
\end{thebibliography}
\end{document}